\documentclass[epj]{webofc}
\usepackage[varg]{txfonts}   
\usepackage{array} 
\usepackage{units}
\usepackage{upgreek}
\usepackage{todonotes}
\wocname{\includegraphics[width=0.25cm,clip]{logo} ARENA-2016}
\woctitle{ARENA-2016}
\renewcommand{\deg}[1]{\ensuremath{#1^{\circ}}\xspace}
\newcommand{\eV}[1]{\ensuremath{\unit[10^{#1}]{eV}}}
\newcommand{\e}[1]{\ensuremath{\cdot 10^{#1}}}
\newcommand{\ece}{\ensuremath{\varepsilon}\xspace}

\newcommand{\eps}{\ensuremath{S_{120}}\xspace}

\newcommand{\epsa}{\ensuremath{S_{120}^{\mathrm{1A}}}\xspace}
\newcommand{\ind}[2]{\ensuremath{#1_{\mathrm{#2}}}}
\newcommand{\dind}[3]{\ensuremath{#1_{\mathrm{#2}}^{\mathrm{#3}}}}
\newcommand{\Sth}{\ensuremath{S_{\mathrm{th}}}\xspace}
\newcommand{\rth}{\ensuremath{r_{\mathrm{th}}}\xspace}

\begin{document}
\title{Tunka-Rex: energy reconstruction with a single antenna station (ARENA2016)}
%
%

\author{\firstname{R.} \lastname{Hiller}\inst{1}\fnsep\thanks{\email{roman.hiller@kit.edu}}
\and
\firstname{P.~A.} \lastname{Bezyazeekov}\inst{2}
\and
\firstname{N.~M.} \lastname{Budnev}\inst{2}
\and
\firstname{O.} \lastname{Fedorov}\inst{2}
\and
\firstname{O.~A.} \lastname{Gress}\inst{2}
\and
\firstname{A.} \lastname{Haungs}\inst{1}
\and
\firstname{T.} \lastname{Huege}\inst{1}
\and
\firstname{Y.} \lastname{Kazarina}\inst{2}
\and
\firstname{M.} \lastname{Kleifges}\inst{3}
\and
\firstname{E.~E.} \lastname{Korosteleva}\inst{4}
\and
\firstname{D.} \lastname{Kostunin}\inst{1}
\and
\firstname{O.} \lastname{Kr\"omer}\inst{3}
\and
\firstname{V.} \lastname{Kungel}\inst{1}
\and
\firstname{L.~A.} \lastname{Kuzmichev}\inst{4}
\and
\firstname{N.} \lastname{Lubsandorzhiev}\inst{4}
\and
\firstname{R.~R.} \lastname{Mirgazov}\inst{2}
\and
\firstname{R.} \lastname{Monkhoev}\inst{2}
\and
\firstname{E.~A.} \lastname{Osipova}\inst{4}
\and
\firstname{A.} \lastname{Pakhorukov}\inst{2}
\and
\firstname{L.} \lastname{Pankov}\inst{2}
\and
\firstname{V.~V.} \lastname{Prosin}\inst{4}
\and
\firstname{G.~I.} \lastname{Rubtsov}\inst{5}
\and
\firstname{F.~G.} \lastname{Schr\"oder}\inst{1}
\and
\firstname{R.} \lastname{Wischnewski}\inst{6}
\and
\firstname{A.} \lastname{Zagorodnikov}\inst{2}
~(Tunka-Rex Collaboration) 
}

\institute{
Institut f\"ur Kernphysik, Karlsruhe Institute of Technology (KIT), Karlsruhe, Germany  
\and
Institute of Applied Physics, Irkutsk State University (ISU), Irkutsk, Russia  
\and
Institut f\"ur Prozessdatenverarbeitung und Elektronik, Karlsruhe Institute of Technology (KIT), Germany
\and
Skobeltsyn Institute of Nuclear Physics, Lomonossov University (MSU), Moscow, Russia
\and
Institute for Nuclear Research of the Russian Academy of Sciences, Moscow, Russia  
\and
Deutsches Elektronen-Synchrotron (DESY), Zeuthen, Germany
}

\abstract{
The Tunka-Radio extension (Tunka-Rex) is a radio detector for air showers in Siberia.
From 2012 to 2014, Tunka-Rex operated exclusively together with its host experiment, the air-Cherenkov array Tunka-133, which provided trigger, 
data acquisition, and an independent air-shower reconstruction.
It was shown that the air-shower energy can be reconstructed by Tunka-Rex with a precision of 15\% for events with signal in at least 3 antennas, 
using the radio amplitude at a distance of 120\,m from the shower axis as an energy estimator.
Using the reconstruction from the host experiment Tunka-133 for the air-shower geometry (shower core and direction),
the energy estimator can in principle already be obtained with measurements from a single antenna, close to the reference distance.
We present a method for event selection and energy reconstruction, requiring only one antenna, and achieving a precision of about 20\%.
This method increases the effective detector area and lowers thresholds for zenith angle and energy, resulting in three times more events than in the standard reconstruction.
}
\maketitle
\section{Introduction}
\label{sec_intro}
After almost being forgotten for about 30 years, the radio detection technique 
resurfaced during the last decade due to the advances in digital technology~\cite{SchroederReview2016,HuegeReview2016}.
It has beneficial properties for the energy measurement, 
like low systematic uncertainties from hadronic processes in the air-shower development, and the possibility to measure the signal on an absolute scale~\cite{TunkaRexScale2016}.
In recent years, sophisticated methods were developed by the latest generation 
of radio experiments to optimize the energy reconstruction~\cite{2014ApelLOPES_MassComposition,TunkaRex_Xmax2016,KostuninTheory2015,AERAenergyPRL2015}.
One of the current radio experiments is Tunka-Rex, the radio extension of Tunka-133~\cite{TunkaRex_NIM_2015,Schroeder_ARENA2016,Kostunin_ARENA2016}.
Tunka-Rex is devoted to demonstrating the competitiveness of an economic radio detector. 
In this work we present a novel method for reconstructing the air-shower energy with a single antenna station.
This optimizes the efficiency for energy reconstruction and allows for a minimal density of radio antennas. 
Details on the overview given here can be found inf Ref.~\cite{Hiller2016}.

\section{Single antenna energy reconstruction}
\label{sec_method}
The Tunka-Rex energy estimator \eps represents the reconstructed electric field, 
corrected for azimuthal asymmetry and geomagnetic angle, at a distance of \unit[120]{m} from the shower axis~\cite{TunkaRex_Xmax2016}.
If there is an antenna close to this distance, \eps can be estimated by this single 
measurement already.
The systematic uncertainty on this measurement certainly increases with
higher distance from the shower axis, and a question answered here is at which distance this uncertainty becomes the limiting factor. 
Using a single antenna to determine the energy of an air shower minimizes the threshold on energy and zenith angle
and increases the effective detector area, maximizing the capabilities of the radio part of a hybrid detector for air showers.

As a single antenna has no sensitivity to fine features of the lateral distribution function (LDF) of 
radio amplitudes, we use the simplest model for the LDF, the exponential function,
to extrapolate the signal in a single antenna to a distance of \unit[120]{m} from the shower axis
\begin{equation}
  \epsa = \dind{S}{meas}{1A} \cdot \exp( \eta \cdot (r-\unit[120]{m})) .
  \label{eq:simpleexp}
\end{equation}
$\eta$ is the slope and $\dind{S}{meas}{1A}$ is the amplitude of the reconstructed electric field, corrected for azimuthal asymmetry, 
of the antenna at distance $r$ from the shower axis.
Although being the simplest among the common models for the LDF, it was demonstrated 
that a resolution of 15\%-20\% can be reached with this model, 
when taking the average azimuthal asymmetry of the signal into account~\cite{Hiller2016}.
No other, more complicated model so far shows significantly better resolution~\cite{2014ApelLOPES_MassComposition,TunkaRex_Xmax2016,AERAenergyPRL2015}.

The shower core and direction, which are also required to apply Eq.~\ref{eq:simpleexp}, are reconstructed from the other component of the hybrid detector,
which for Tunka-Rex is Tunka-133. 
For the slope $\eta$, we use the following parametrization with the zenith angle $\theta$ and $\eta_0=\unit[8.2\e{-3}]{m^{-1}}$:
\begin{equation}
\eta=\eta_0\cdot\cos\theta,
\end{equation}
where  $\eta$ varies from event to event about $\sigma_{\eta}=\unit[3\e{-3}]{m^{-1}}$ in Tunka-Rex,
which does not limit the accuracy of the energy determination with a single antenna for most events.

In case multiple antennas with signal are available in an event, for consistency, only the one with the
smallest estimated uncertainty is used, where with $\Delta r=r-\unit[120]{m}$ the uncertainty is estimated to
\begin{equation}
\label{eq:1antennaUC}
  \ind{\sigma}{tot}=\exp(\eta \cdot \Delta r)\cdot\sqrt{\ind{\sigma}{m}^2 + 
  \left(\dind{S}{meas}{1A}\cdot\Delta r \cdot\sigma_{\eta}\right)^2},
\end{equation} 
The two contribution in the uncertainty arise due to noise $(\ind{\sigma}{m})$ and shower-to-shower fluctuations of the slope
and both increase with higher extrapolated distances.

While reducing the minimum number of antennas from 3 to 1 lowers the threshold for air-shower detection,
it also increases the risk of false positive events due to background.
There are several possibilities to circumvent this, starting by simply 
increasing the signal-to-noise threshold.
However, as the analysis is tailored for radio detectors in hybrid mode anyway, 
we decided to use the full information available from Tunka-133.
Therefore, a model for the footprint of radio amplitudes was developed, and for each Tunka-133 event the distance \rth from the shower axis is calculated, at which the radio amplitude is expected to pass the detection threshold.
For a detailed description of this model we refer to Ref.\cite{Hiller2016} and give only the formula here:
\begin{equation}
\label{eq:rth}
  \rth(E,\theta,\alpha)=r_0 + \frac{1}{\eta_0\cos\theta}\ln \frac{E\cdot \sqrt{\sin^2\alpha+\ece^2}}{\kappa \Sth}
\end{equation}
$\alpha$ is the geomagnetic angle and the remaining parameters and their meaning are given in Tab.~\ref{tab:effmodel}. 
Only antennas within a distance from the shower axis smaller than \rth are considered for energy reconstruction.
The energy, geomagnetic angle and zenith angle needed to calculate \rth are again determined by Tunka-133. 
However, unlike the shower geometry, the energy reconstruction of Tunka-133 is exclusively used for the purpose of selecting events (and antennas) entering the radio analysis.
\begin{table}[htb]
\caption{
Parameters for the footprint of the radio signal above threshold from Eq.~\ref{eq:rth} for Tunka-Rex.
They parameters were partly measured and partly determined from CoREAS simulations \cite{Hiller2016}.
}
\centering
\begin{tabular}{l l l}

\hline
Parameter & Value & Origin\\
\hline
$r_0$ reference distance    & \unit[120]{m} & Simulation\\ 
$\eta_0$ LDF slope & \unit[8.2\e{-3}]{m$^{-1}$} & Measurement\\
$\ece\,$ charge-excess contribution  & 0.085 & Simulation\\  
$\kappa\,$ radio signal strength & \unit[884]{$\mathrm{\frac{EeV}{V/m}}$} & Simulation\\
$\Sth$ detection threshold   & \unit[90]{$\upmu$V/m} & Measurement
\end{tabular}
\label{tab:effmodel}
\end{table}

\section{Performance on data of first two seasons}
\label{sec_performance}
To test the performance of the single antenna energy reconstruction we use data from the 
first and second season of Tunka-Rex (Oct. 2012 - April 2014) and look at events with energies reconstructed by Tunka-133 above 
\eV{16.5} and zenith angles below \deg{50}.
After removing background from the trace, the electric field is reconstructed using the 
incoming direction reconstructed by Tunka-133.
Then, the signal amplitude $S$ is determined as the peak of a Hilbert envelope in the signal window and 
the noise $N$ as the RMS of a noise window shortly before the signal window \cite{TunkaRex_NIM_2015}.
Only antennas with a signal-to-noise ratio (SNR = $S^2/N^2$) above 10 and within \rth are further considered, i.e.,
both the measured SNR and the SNR expected from the footprint model are required to be above 10. 

\begin{figure}[tb]
\includegraphics[height=0.26\textheight]{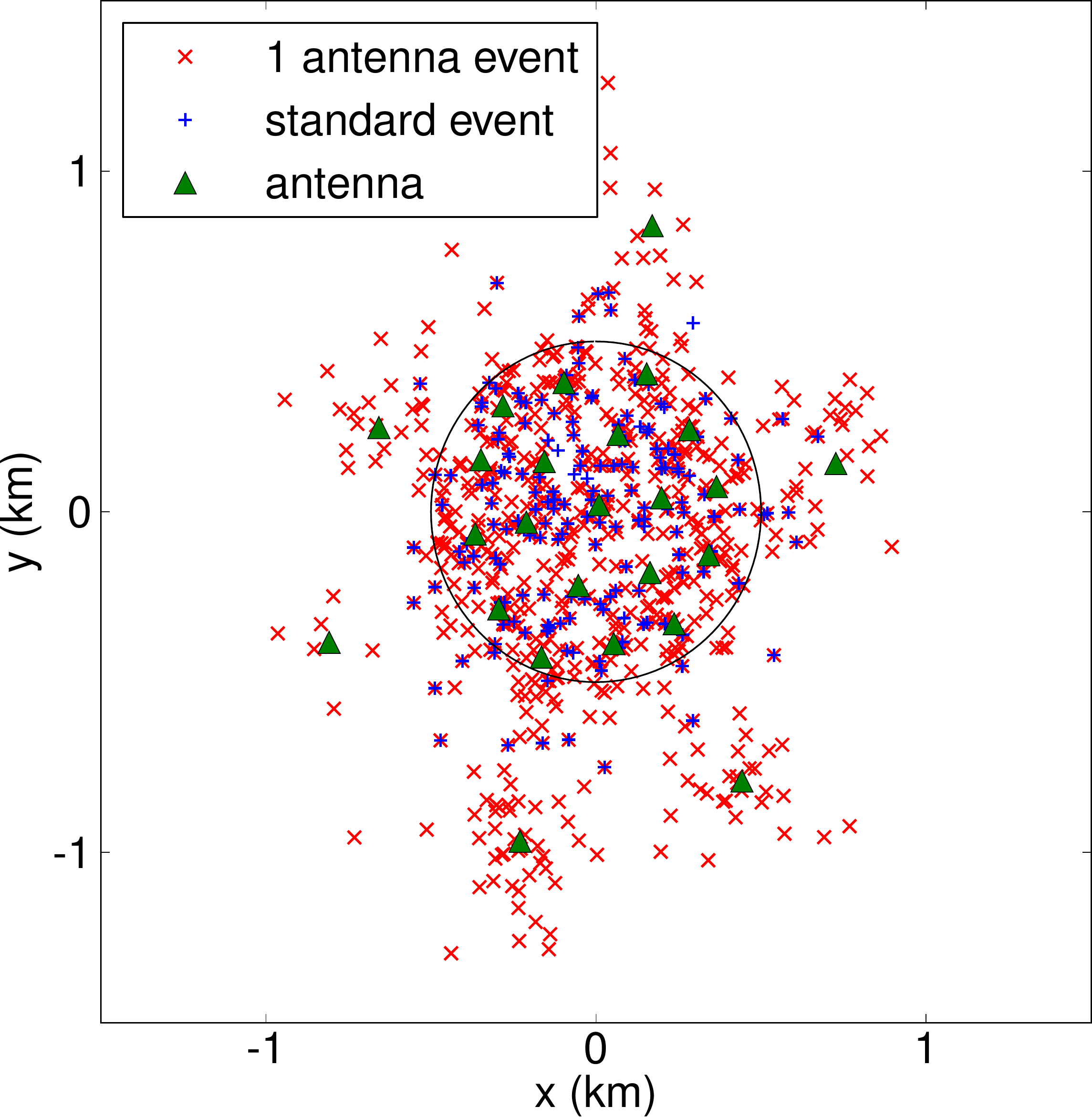}
\hfill
\includegraphics[height=0.26\textheight]{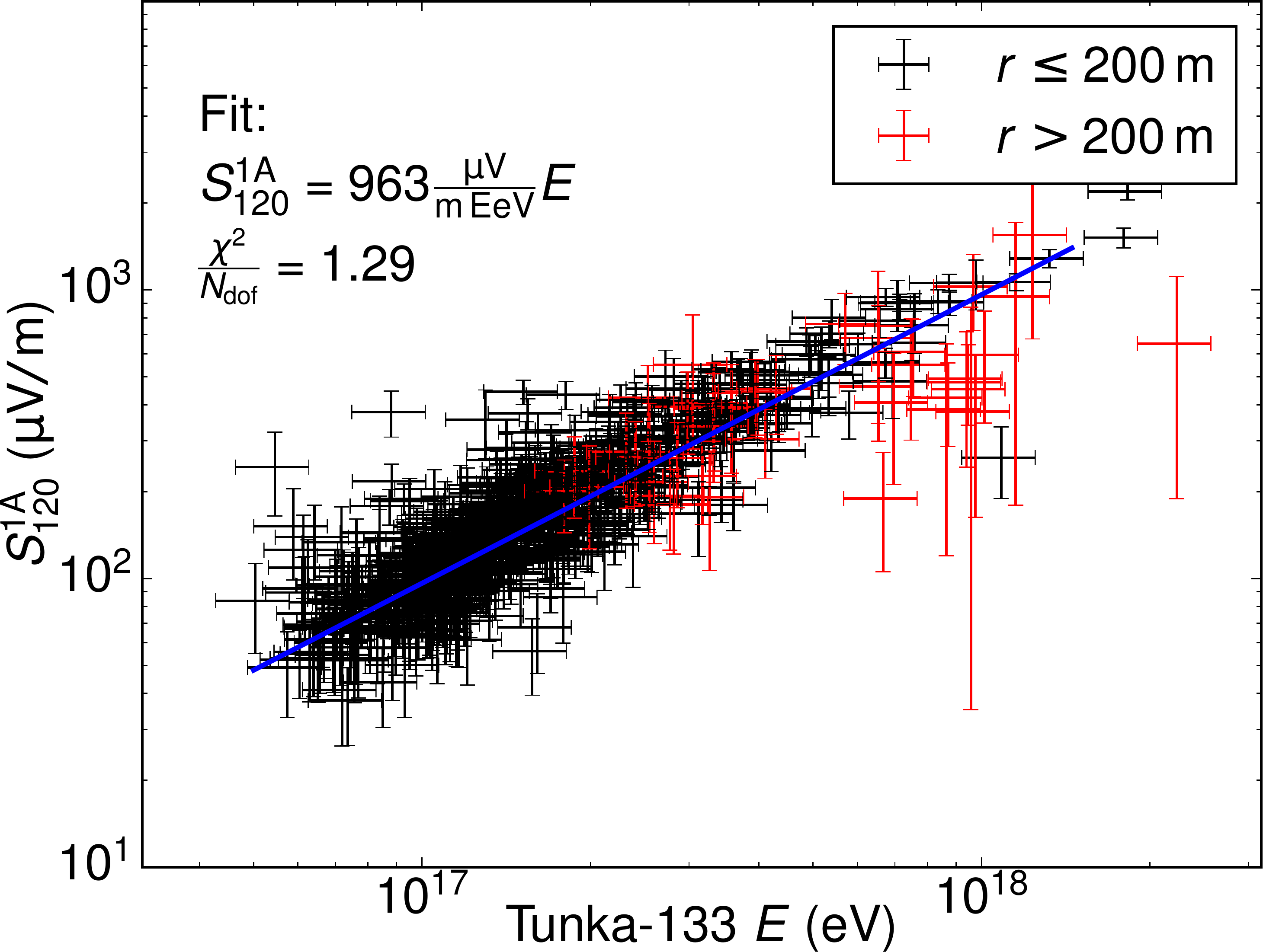}
\caption{{\bf Left}: Core distribution of events from standard and single-antenna analyses.
Events of the latter extend beyond the borders of the inner, dense array.
{\bf Right}: The energy estimator of the singe-antenna analysis \epsa versus Tunka-133 energy.
Events far from the shower axis suffer from high systematic uncertainties due to the extrapolation of the LDF. 
}
\label{fig-1}       
\end{figure}

649 events pass these requirements, more than three times that in the standard analysis, requiring three antennas with signal~\cite{TunkaRex_NIM_2015}.
In Fig.~\ref{fig-1} (left), the distribution of shower cores of this selection and of the standard reconstruction are shown.
The highest gains in event number are from the outer clusters, where the antenna density of Tunka-Rex decreases dramatically.
Furthermore, the energy threshold decreases from $\eV{16.9}$ to $\eV{16.7}$ and many events with zenith angles below \deg{30} pass the cuts, 
which in the standard analysis only happens for the highest energies.
Thus, dropping the requirement for 3 antennas increases the event number at low and high energies, by increasing the efficiency and
effective detector area.

In Fig.~\ref{fig-1} (right), the single-antenna estimator for the energy \epsa is shown versus the Tunka-133 energy.
While \epsa shows a clear correlation with the energy, there are a few outliers, almost exclusively from antennas further than \unit[200]{m}
from the shower axis, where systematic uncertainties from the LDF parametrization start to dominate.
Removing these events, 606 events are left. 
The resolution obtained from the spread of the relative difference between the two energy estimators is 25\%.
Assuming no correlation between them and 15\% uncertainty for the energy from Tunka-133~\cite{Tunka133_NIM2014},
a resolution of 20\% is obtained for \epsa.

\section{Conclusion}
We presented a method for the reconstruction of the air-shower energy with the radio part of a hybrid detector
using data from a single antenna station only.
With Tunka-Rex an energy resolution of 20\% can be reached with this method, obtaining more than 3 times
more events than with the standard selection requiring three antennas.
Minimizing the required antenna density, this approach achieves the maximum efficiency possible 
with an economic radio extension, reaching down to the limit of amplitude based methods:
the sensitivity of the individual antennas.

\section*{Acknowledgments}
\footnotesize
The construction of Tunka-Rex was funded by the German Helmholtz association and the Russian Foundation for Basic Research (grant HRJRG-303). 
Moreover, this work has been supported by the Helmholtz Alliance for Astroparticle Physics (HAP), by Deutsche Forschungsgemeinschaft (DFG) grant SCHR 1480/1-1, and by the Russian grant RSF 15-12-20022. 
\bibliography{arena2016}

\end{document}